\documentclass[twocolumn,showpacs,amsmath,amssymb]{revtex4}
\usepackage{graphicx}
\usepackage{dcolumn}
\usepackage{bm}
\preprint{APS/123-QED}
\begin{document}

\title{Short-length storage of intense optical pulses in solid
by adiabatic passage}
\author{G. G. Grigoryan}
\email{gaya@ipr.sci.am}
 \affiliation{Institute for Physical Research, 378410, Ashtarak-2, Armenia}
\author{Y. T. Pashayan-Leroy}
 \affiliation{Institut Carnot de
Bourgogne, UMR 5209 CNRS - Universit\'e de Bourgogne, BP 47870,
21078 Dijon, France}
\author{C. Leroy}
 \affiliation{Institut Carnot de
Bourgogne, UMR 5209 CNRS - Universit\'e de Bourgogne, BP 47870,
21078 Dijon, France}
\author{S.~Gu\'erin}
 \affiliation{Institut Carnot de
Bourgogne, UMR 5209 CNRS - Universit\'e de Bourgogne, BP 47870,
21078 Dijon, France}

\date{\today}

\begin{abstract}
We propose a novel scheme of storage of intense pulses which
allows a significant reduction of the storage length with respect
to standard schemes. This scheme is particularly adapted to store
optical information in media with fast relaxations.
\end{abstract}

 \maketitle


\section{Introduction}
The propagation of two pulses in resonant interaction in a
$\Lambda$- atomic medium has been widely studied during the last
decade, in particular with application to quantum information
processing (see, e.g., reviews \cite{Lukin2003}). The possibility
of quantum information storage in the gas phase has been
demonstrated in cold atoms \cite{Liu} and in hot ensembles
\cite{Lukin2001}. Achieving information storage in solid-state
systems, which are more attractive due to their higher density,
compactness, and absence of diffusion, has been pursued
\cite{Kuznetsova}. The main drawbacks preventing the efficiency of
storage for such solid-state materials are huge inhomogeneous
broadenings and high rates of decoherence. For instance, in
crystalline films doped with rare-earth metals the rates of
transverse relaxations amount to tens of GHz \cite{Johnsson}. In
practice an efficient storage of information requires rather large
optical length such that even a weak loss rate will ruin it
\cite{Novikova}. In order to reduce the inhomogeneous broadening,
it was proposed in a number of works to use the so-called hole
burning technique \cite{Shahriar}. It has been shown that media
prepared in such a way allow efficient coherent population
transfer in $\Lambda$ systems \cite{Goto}. However, the hole
burning technique faces a so far unsolved problem that leads to
the reduction of the optical length of the samples, which will be
detrimental in general for the efficiency of the storage.

In the present work we show that, for the storage of optical
information (i.e. of classical fields), the recording length can
be dramatically reduced with the use of intense pulses. The
possibility to store and retrieve optical information in resonant
media has mainly been studied in the ``linear approximation'' with
respect to the so-called probe field that has to be stored, i.e.
for a weak probe pulse. It was usually assumed that the control
pulse propagates in a medium without pulse shape change.
Numerical simulation of this problem
without restrictions on the probe intensity was performed in
\cite{Dey}. Analytical studies of the problem taking into account
the group velocities of both pulses were performed in
\cite{Grigoryan} in the limit of pulses of duration much shorter
than all the relaxation times,  but sufficiently long to allow
adiabatic evolution during the interaction. It has been found that
the length of information storage depends remarkably on the ratio
between the oscillator strengths of the adjacent transitions that
determine the group velocities of the pulses in a medium. The
present work aims at a complete analytical study of information
storage in case of arbitrary relaxation times and intensities of
the probe and control fields. We show analytically and numerically
that it is possible, for a proper choice of solid-state medium, to
dramatically reduce the optical length needed to store intense
short pulses. This would in particular make more efficient the use
of the hole burning technique.

The paper is organized as follows: We first describe the model,
and analyze the propagation in the adiabatic limit. We next derive
the conditions of storage for short lengths of the medium, before
concluding.

\begin{figure}[hbtp]
 \centering
 \includegraphics {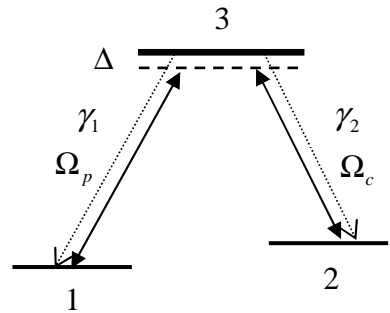}
 \caption{Level scheme diagram for a $\Lambda$-system interacting
 with two laser pulses.}
 \end{figure}

\section{The model}

We consider a medium of $\Lambda$ three-level atoms with two
ground states $|1\rangle,|2\rangle$ and a metastable excited state
$|3\rangle$, of energy $\hbar\omega_i$, $i=1,2,3$, and two laser
pulses referred to as the probe and control pulses, coupling the
transition respectively $1\leftrightarrow 3$ and $2\leftrightarrow
3$ (see Fig. 1). The interaction Hamiltonian in the rotating wave
approximation can be written in the basis
$\{|1\rangle,|2\rangle,|3\rangle\}$ as
\begin{displaymath}
    H=\frac{\hbar }{2}\left[
    \begin{array}{ccc}
        0 & 0 & -\Omega_{p}^{*} \\
        0 & 2\Delta & -\Omega_{c}^{*}\\
        -\Omega_{p} & -\Omega_{c} & 0
    \end{array}\right]
\end{displaymath}
with
$\Delta=\omega_3-\omega_1-\omega_p=\omega_2-\omega_1-\omega_p$ the
one-photon detuning, $\omega_{p}$ ($\omega_{c}$) the probe
(coupling) laser frequency, $\Omega_{p,c}={\cal
E}_{p,c}\mu_{p,c}/\hbar$ the Rabi frequencies associated to the
corresponding field amplitudes ${\cal E}_{p,c}$ and to the dipole
moments $\mu_{p,c}$ of the corresponding atomic transitions. We
consider an exact two-photon resonance.

The propagation of two laser pulses in the medium is described by
the Maxwell equations which, in the slowly-varying-amplitude
approximation and in the moving coordinate system
\begin{equation}
\label{moving} \eta=x,\quad \tau=t-x/c
\end{equation}
(such that the original wave operator $\partial/\partial
t+c\partial/\partial x$ becomes $c\partial/\partial \eta$), read
\cite{Lukin2003}
\begin{equation}
\label{Maxwell} \frac{\partial\Omega_p}{\partial
\eta}=iq_p\rho_{31},\quad \frac{\partial\Omega_c}{\partial
\eta}=iq_c\rho_{32}.
\end{equation}
Here $q_{p,c}=2\pi\omega_{p,c}|\mu_{p,c}|^2N/\hbar c$ are the
coupling factors with $N$ the density of atoms.  The density
matrix elements $\rho_{ij}$ are determined by the system of
equations (where the dot denotes the derivative $\partial/\partial
\tau$, since $\partial/\partial t=\partial/\partial\tau$)
\begin{subequations}
\label{syst}
\begin{eqnarray}
\dot\rho_{11}&=&\gamma_1\rho_{33}-2\text{Im}(\Omega_p^{\ast}\rho_{31})\\
\dot\rho_{22}&=&\gamma_2\rho_{33}-2\text{Im}(\Omega_c^{\ast}\rho_{32})\\
\dot\rho_{33}&=&-(\gamma_1+\gamma_2)\rho_{33}
+2\text{Im}(\Omega_p^{\ast}\rho_{31}+\Omega_c^{\ast}\rho_{32})\\
\dot\rho_{31}&=&-(\Gamma+i\Delta)\rho_{31}+i\Omega_p(\rho_{11}-\rho_{33})
+i\Omega_c\rho_{21}\\
\dot\rho_{32}&=&-(\Gamma+i\Delta)\rho_{32}+i\Omega_c(\rho_{22}-\rho_{33})
+i\Omega_p\rho_{21}^{\ast}\\
\dot\rho_{21}&=&-\gamma_{c}\rho_{21}-i\Omega_p\rho_{32}^{\ast}
+i\Omega_c^{\ast}\rho_{31}
\end{eqnarray}
\end{subequations}
with $\gamma_1$ and $\gamma_2$ the rates of spontaneous emission
from state 3 to  states 1 and 2 respectively, $\gamma_c$ the
dephasing rate between the two ground states, and $\Gamma$ the
transverse relaxation rate. We moreover assume pulses of durations
much shorter than the characteristic time of decoherence of the
ground states (usually microseconds in solids) such that
$\gamma_c$ will be neglected. The envelopes of the fields ${\cal
E}_{p,c}\equiv{\cal E}_{p,c}(\eta,\tau)$, the Rabi frequencies
$\Omega_{p,c}\equiv\Omega_{p,c}(\eta,\tau)$, and the density
matrix components $\rho_{ij}\equiv\rho_{ij}(\eta,\tau)$ depend on
the position $\eta=x\in[0,L]$ in the medium of length $L$ and on
the time $\tau=t-x/c$. The boundary conditions for the Maxwell
equations are the pulses ${\cal E}_{p}(0,\tau)\equiv {\cal
E}_{p,0}(\tau)$, ${\cal E}_{c}(0,\tau)\equiv {\cal E}_{c,0}(\tau)$
given at the entrance $\eta=x=0$ of the medium. We assume that all
atoms in the medium are in state 1 before the interaction with the
laser fields.

\section{Adiabatic propagation}
It is well known that the use of a counterintuitive sequence of
pulses in two-photon resonance, here the coupling pulse switched
on before the probe pulse, leads, in the limit of slow evolution
of the pulses, to an adiabatic passage along the dark state (of null
eigenvalue) immune to decoherence. The system of linear equations
(\ref{syst}) can be recast as $\frac{d}{ds}\rho(\epsilon
s)=L(\epsilon s)\rho(\epsilon s)$ with $L$ being the associated linear
operator and $\epsilon$ the formal (small) adiabatic parameter (we
have omitted here the dependence on $\eta$ for simplification).
The change of variable $\tau=\epsilon s$ leads to
\begin{equation}
\label{Lindblad_adiab} \epsilon
\frac{d}{d\tau}\rho(\tau)=L(\tau)\rho(\tau).
\end{equation}
 The components of the eigenvector
$\rho^{(0)}(\tau)$ of null eigenvalue satisfying
$0=L(\tau)\rho^{(0)}(\tau)$ read
\begin{subequations}
\label{adiab}
\begin{eqnarray}
\rho^{(0)}_{11}&=&\cos^2\theta,\quad\rho^{(0)}_{22}=\sin^2\theta,\quad
\rho^{(0)}_{33}=0\\
\rho^{(0)}_{21}&=&-e^{i(\varphi_p-\varphi_c)}\sin\theta\cos\theta,\quad
\rho^{(0)}_{31}=\rho^{(0)}_{32}=0
\end{eqnarray}
\end{subequations}
with $\theta\equiv\theta(\eta,\tau)$ as
\begin{equation}
\tan\theta=\frac{|\Omega_p|}{|\Omega_c|},\quad
\Omega_{p,c}=|\Omega_{p,c}|e^{i\varphi_{p,c}}.
\end{equation}
In the adiabatic limit, there is thus no coherence between the
ground states and the excited state, which induce no pulse
distortion during the propagation \cite{Lukin2003}.

To determine non-adiabatic corrections, we make around the
adiabatic solution $\rho(\tau)=\rho^{(0)}(\tau)$ the
superadiabatic expansion:
\begin{equation}
\label{expansion} \rho=\rho_0^{(0)}+\epsilon\rho_0^{(1)}+
\epsilon^2\rho_0^{(2)}+\cdots
\end{equation}
Inserting this expansion in Eq. (\ref{Lindblad_adiab}) and
identifying the orders of $\epsilon$, we obtain a system of linear
equations that can be solved iteratively. For the first order
\begin{equation}
\dot\rho_0^{(0)}=L\rho_0^{(1)},
\end{equation}
denoting $\varphi=\varphi_p-\varphi_c$, we get
\begin{subequations}
\label{superadiab}
\begin{eqnarray}
\label{superadiaba}
\rho^{(1)}_{11}&=&\frac{\sin2\theta}{\Omega^2}\left(\Gamma\dot\theta
-\frac{\Delta\dot\varphi}{2} \sin2\theta\right),\\
\rho^{(1)}_{22}&=&-\rho^{(1)}_{11},\quad \rho^{(1)}_{33}=0\\
\label{superadiabb}
\rho^{(1)}_{21}&=&\frac{e^{i\varphi}}{\Omega^2}\Bigl[\dot\theta
(\Gamma\cos2\theta+i\Delta)\nonumber\\
&&\qquad\qquad+\frac{\dot\varphi}{2}\sin2\theta
(-\Delta\cos2\theta+i\Gamma)\Bigr],\\
\label{superadiabc}
\rho^{(1)}_{31}&=&e^{i\varphi_p}\frac{\cos\theta}{\Omega}\left(i\dot\theta
-\frac{\dot\varphi}{2}\sin2\theta\right),\\
\label{superadiabd}
\rho^{(1)}_{32}&=&e^{i\varphi_c}\frac{\sin\theta}{\Omega}\left(-i\dot\theta
+\frac{\dot\varphi}{2}\sin2\theta\right)
\end{eqnarray}
\end{subequations}
 with
\begin{equation}
\Omega^2\equiv \Omega^2(\eta,\tau)=|\Omega_p|^2+|\Omega_c|^2.
\end{equation}
Equations (\ref{superadiaba}) and (\ref{superadiabb}) show that at
the first superadiabatic order the excited state stays unpopulated
and that the adiabatic dynamics along the dark state is subjected
to a loss of order $\Gamma/\Omega^2T$ and to a pure dephasing loss
of order $\Delta/\Omega^2T$ with $T$ the characteristic time of
interaction. Equations (\ref{superadiabc}) and (\ref{superadiabd})
show the transient appearance of coherences between the ground
states and the excited state, which depend on the first-order
non-adiabatic loss $\dot\theta$ (and also on $\dot\varphi$), but
\textit{not} on the relaxation loss. These coherences will be
responsible of pulse distortion during the propagation [through
the Maxwell equations (\ref{Maxwell})], which will be thus
independent of the relaxation loss. Inspection of the solution
(\ref{superadiab}) allows one to identify the small parameter
$\epsilon\ll1$ that permits the expansion (\ref{expansion}) (at
least at the first order). We get the conditions:
\begin{equation}
\label{conds} \Omega T\gg 1,\quad\frac{\Omega^2 T}{\Delta}\gg 1,
\quad\frac{\Omega^2 T}{\Gamma}\gg 1,\quad\frac{\Omega^2
T}{\gamma_c}\gg 1.
\end{equation}
The two first conditions (large pulse area and small one-photon
detuning) are the standard condition for adiabatic passage for a
non-lossy system. The third condition can be interpreted as the
requirement of a long interaction time and a large pulse area with
respect to the decay of the dynamics along the dark state. This
third condition shows that a fast relaxation $\Gamma T\gg1$
usually encounters in solids can be compensated by a strong pulse
area satisfying
\begin{equation}
\label{cond_fast_relax} (\Omega T)^2\gg \Gamma T.
\end{equation}
We have added in Eqs. (\ref{conds}) the well known fourth
condition considering an additional decoherence as a pure
dephasing of rate $\gamma_c$ between the two ground states.

It is frequently mentioned that the first condition of
adiabaticity is not required and that the control pulse may have a
rectangular shape (see for instance \cite{Matsko}). It is indeed
shown in this Ref. \cite{Matsko} that in case of a rectangular
control pulse an efficient storage of information is possible.
However, the parameter $T$ of the above conditions characterizes
the time of interaction when the probe and the control pulses
\textit{both} interact, i.e. when they \textit{overlap}. One can
thus consider a rectangular control pulse but turned on earlier
and turned off later than the probe pulse if one preserves the
adiabaticity during the field overlapping.

Using the density matrix at first order (\ref{expansion}) with
(\ref{adiab}) and (\ref{superadiab}), we can rewrite the Maxwell
equations (\ref{Maxwell}) as
\begin{subequations}
\label{Maxwell2}
\begin{eqnarray}
\label{Maxwell2a} \frac{\partial\theta}{\partial
\eta}+\frac{1}{u}\frac{\partial\theta}{\partial
\tau}&=&0,\\
\label{Maxwell2b} \frac{\partial}{\partial
\eta}\left(\frac{\Omega^2}{Q}\right)&=&0,\\
\label{Maxwell2c} \frac{\partial\varphi}{\partial
\eta}+\frac{1}{u}\frac{\partial\varphi}{\partial \tau}&=&0
\end{eqnarray}
\end{subequations}
with
\begin{subequations}
\begin{eqnarray}
\label{Q} Q&\equiv&
Q(\eta,\tau)=\frac{q_pq_c}{q_c\sin^2\theta(\eta,\tau)
+q_p\cos^2\theta(\eta,\tau)},\\
\label{u} u&\equiv&
u(\eta,\tau)=\frac{Q(\eta,\tau)\Omega^2(\eta,\tau)}{q_pq_c},
\end{eqnarray}
\end{subequations}
respectively the two-photon transition strength and the
group velocity, and the boundary condition
\begin{equation}
\label{theta_bound} \theta(0,\tau)\equiv \theta_0(\tau).
\end{equation}
 Denoting the photon fluxes ($j=p,c$)
\begin{subequations}
\begin{eqnarray}
n_{j}&\equiv& n_{j}(\eta,\tau)=\frac{c|{\cal
E}_{j}(\eta,\tau)|^2}{2\pi\hbar\omega_{j}}=\frac{
N|{\Omega}_{j}(\eta,\tau)|^2}{q_{j}},\\
n&\equiv& n(\eta,\tau)=n_p(\eta,\tau)+n_c(\eta,\tau),
\end{eqnarray}
\end{subequations}
Eq. (\ref{Maxwell2b}) can be rewritten as $\partial n/\partial
\eta=0$ and interpreted as the conservation of the photon fluxes
during the propagation:
 \begin{equation}
\label{conserv} n(\eta,\tau)=n(0,\tau)\equiv n_0(\tau) \equiv
N\Bigl(\frac{|{\Omega}_{p,0}(\tau)|^2}{q_{p}}+\frac{
|{\Omega}_{c,0}(\tau)|^2}{q_{c}}\Bigr).
\end{equation}
Note that this holds only in the limit of no population in the
excited state (i.e. at the first order of the superadiabatic
expansion) \cite{Grigoryan}.

The obtained system of equations coincides with that studied in
detail in literature, where pulses of durations much shorter than
all relaxation times were considered (see e.g. \cite{Grigoryan}).
However, the system of equations obtained in previous papers (e.g.
in \cite{Grobe}) is valid under condition of interaction
adiabaticity $\Omega T\gg 1$, whereas at high relaxation rates the
inequality (\ref{cond_fast_relax}) should be fulfilled. Solutions
of the system of equations (\ref{Maxwell2}) have been obtained and
studied for different regimes of propagation in a number of works
(see, e.g., the review \cite{Lukin2003}). The method of
characteristics, recalled in Appendix \ref{app_characteristics},
allows one to determine them.

 In particular, equal transition strengths
$q_p=q_c$ lead to a constant value $Q=q_p$, and from Eq.
(\ref{Maxwell2b}) to $\Omega$ independent of $\eta$, thus of shape
invariance during the propagation:
$\Omega(\eta,\tau)\equiv\Omega_0(t-x/c)$, as well as $u$:
$u(\eta,\tau)\equiv u_0(t-x/c)=\Omega_0^2(t-x/c)/q_p$ with
$\Omega_0(t)$ given at the entrance of the medium. The pulses
[through their Rabi frequencies
$\Omega_p=e^{i\varphi_p}\Omega_0(t-x/c)\sin\theta$ and
$\Omega_c=e^{i\varphi_c}\Omega_0(t-x/c)\cos\theta$] propagate in a
complicated manner through $\Omega_0$ and the mixing angle of
solution [see Eq. (\ref{ODE_sol})]
\begin{equation}
\label{theta} \theta(\eta,\tau)=\theta(0,\xi) \equiv \theta_0(\xi)
\end{equation}
where $\xi\equiv\xi(x,t)$ is defined by (\ref{xi}) (where $t$
should be replaced by $\tau=t-x/c$) \cite{Grobe}
\begin{equation}
\int_{\xi}^{t-x/c}d\tau\,\Omega_0^2(\tau)/q_p=x.
\end{equation}
When $\Omega_0$ is additionally constant, the excitation
propagates at constant velocity satisfying
$1/u=1/c+q_p/\Omega_0^2$. This propagation regime is known as
adiabaton \cite{Grobe}.

 In the more general case of unequal transition
strengths, Eq. (\ref{Maxwell2a}) entails that the mixing angle
still reads \cite{Grigoryan}
\begin{equation}
\label{theta} \theta(\eta,\tau)=\theta(0,\xi)=\theta_0(\xi)
\end{equation}
and from (\ref{Q}) that $Q(\eta,\tau)=Q(0,\xi)\equiv Q_0(\xi)$
with $u(\eta,\tau)=Q_0^2(\xi)n_0(\tau)/Nq_pq_c$, and
$\xi\equiv\xi(x,t)$ determined from Eq. (\ref{xi}) (where $t$
should be replaced by $\tau=t-x/c$):
\begin{equation}
\label{xi2} \int_{\xi}^{t-x/c}d\tau\,n_0(\tau)=\frac{
Nq_pq_c}{Q_0^2(\xi)}x.
\end{equation}
$\Omega^2$ propagates in a more complicated manner than in case of
equal transition strength as
$\Omega^2(\eta,\tau)=Q_0(\xi)n_0(t-x/c)/N$. Since $Q_0$ and $n_0$
do not propagate at the same velocity, $\Omega(\eta,\tau)$ is not
any more shape-invariant during the propagation.

 If we assume constant phases $\varphi_{p,c}$ at the entrance of
the medium, one can conclude from Eq.(\ref{Maxwell2c}) that they
will thus not change during the propagation. Note that this
statement is only valid at the exact two-photon resonance.
Evolution of the phase self-modulation for a two-photon detuning
different from zero has been analyzed in \cite{Grigoryan}.

In the linear approximation (with respect to the probe field),
i.e., in the first order with respect to $\theta$, one has $Q =
q_c$, $\Omega^2=\Omega^2_c$, $u=\Omega_c^2/q_p$. We remark that in
case of equal strengths of adjacent transitions $q_p=q_c$, similar
expressions are obtained with $\Omega^2$ in place of $\Omega_c^2$.
Thus, in case of equal transition strengths the nonlinearity in
the probe field results in replacement of the Rabi frequency of
the control field by the generalized Rabi frequency $\Omega^2$.
The nonlinearity is more important when the adjacent transition
strengths are unequal and can lead to formation of shock-wave
fronts and violation of the interaction adiabaticity, as shown in
\cite{Grigoryan}.

\section{Efficient storage}

The storage of the probe pulse is achieved when its time profile
is mapped into the spatial distribution of the atoms in the
medium. More precisely, the probe pulse can be encoded in the
spatial distribution of the angle $\theta_0(\xi(x,t\to+\infty))$
after the passage of the pulses through the medium, inducing for
instance a storage in the coherence between the lower states (in
the adiabatic limit):
\begin{equation}
\rho_{21}(x,t\to+\infty)\approx-\frac{1}{2}e^{-i\varphi}
\sin[2\theta_0(\xi(x,t\to+\infty))].
\end{equation}
 An efficient
storage requires that the excitation, through $\theta(x,t)$,
propagates adiabatically until it eventually stops, which
necessitates a long enough medium. It is usually achieved for a
strong control pulse switched on before and switched off after a
weak probe. We present below a novel pulse configuration with a
strong probe and a weak control that allows one to shorten
significantly the length of storage.
\begin{figure}[hbtp]
 \centering
 \includegraphics[scale=0.85] {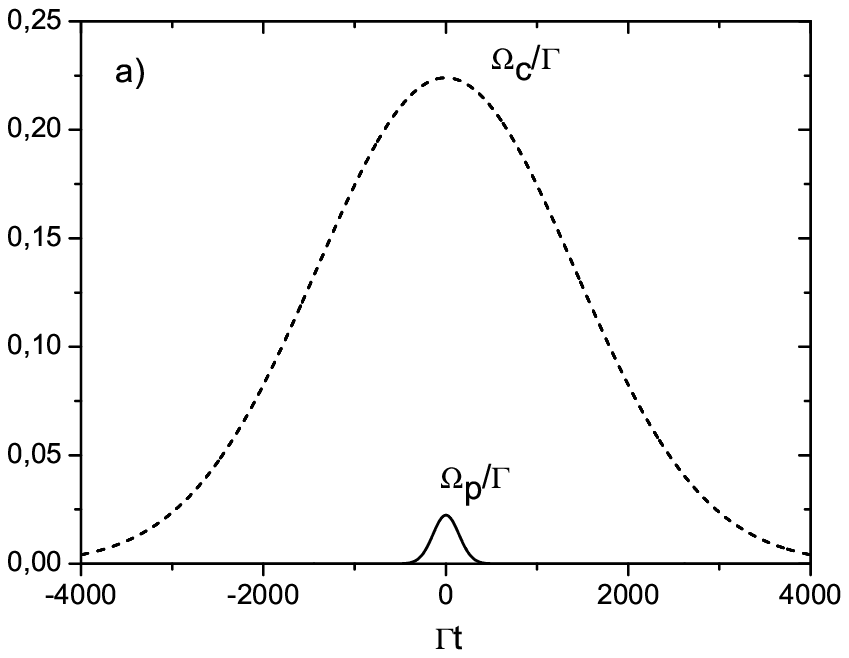}
 \includegraphics[scale=0.85] {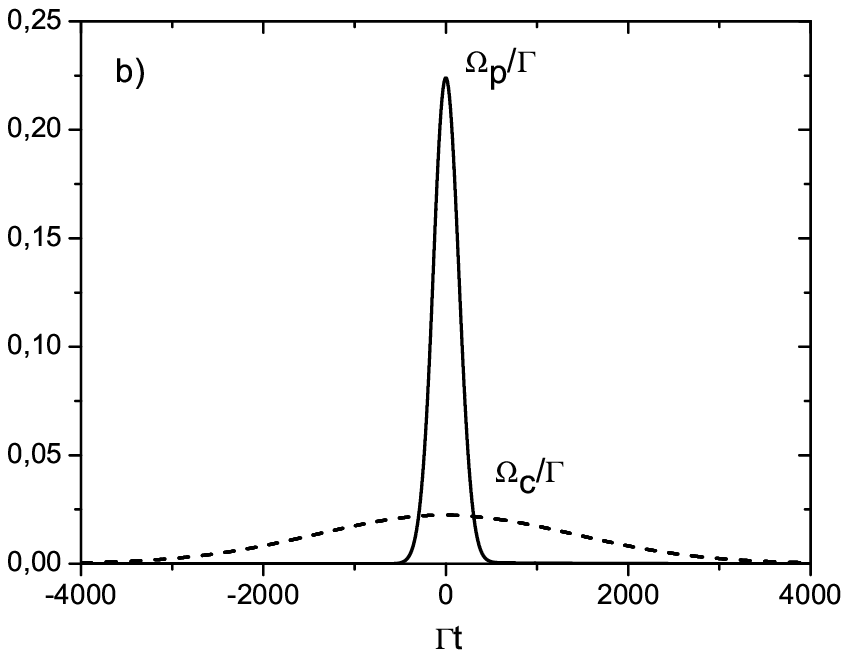}
 \caption{Time envelopes of the control (dashed lines)
 and the probe (full lines) Gaussian pulses
 $\Omega_{j}=\Omega_{j,\text{max}}\exp[-(t/T_{j})^2]$, $j=p,c$,
 at the entrance of the medium with $T_c=10T_p$ and
 $\Gamma T_p=200$: a) Usual scheme (shown here for
 $\Omega_{p,\text{max}}/\Gamma = 2.24\times
10^{-2}$, $\Omega_{c,\text{max}}/\Gamma = 2.24\times 10^{-1}$);
 b) Proposed scheme with $\Omega_{p,\text{max}}/\Gamma = 2.24\times
10^{-1}$, $\Omega_{c,\text{max}}/\Gamma = 2.24\times 10^{-2}$.
}
 \end{figure}
The storage requires the stopping of the excitation, $u(x,t)=0$,
for all the characteristics connected with the initial excitation
at the entrance of the medium $x=0$. We can estimate the length
$x_{\max}$ required to completely store the excitation from
(\ref{xi2}) using the longest characteristic, i.e. $\xi\to-\infty$
and $t\to\infty$ \cite{Grigoryan}:
\begin{subequations}
\label{xmax}
\begin{eqnarray}
x_{\max}&=&\frac{1}{q_p}
\int_{-\infty}^{+\infty}dt\left(q|{\Omega}_{p,0}|^2
+|{\Omega}_{c,0}|^2\right),\ q\equiv\frac{q_c}{q_p}\qquad\\
&\sim& \frac{1}{q_p}\left( q{\Omega}_{p,\max}^2 T_p
+{\Omega}_{c,\max}^2 T_c\right)
\end{eqnarray}
\end{subequations}
with $T_p$ and $T_c$ the duration of respectively the probe and
the control, and $\Omega_{j,\max}\equiv\max_t{\Omega_{j,0}(t)}$,
$j=p,c$. Minimizing this quantity requires thus $q<1$ and a weak
probe pulse, as is well known. On the other hand, adiabaticity of
the interaction is required. Criterion for this adiabaticity has
been analyzed in \cite{Grigoryan}, where it has been shown that,
under the certain condition adiabaticity is broken with formation
of shock-wave fronts for distances exceeding the critical length
$x_0$ estimated as
\begin{equation}
\label{x0}
x_0
\sim
\frac{T_p}{q_p}\frac{\Omega_{p,\max}^2+\Omega_{c,\max}^2}{1-q}.
\end{equation}
The latter approximation holds for a probe pulse of duration $T_p$
shorter than the control.
 This shows that reducing $q$ to lower $x_{\max}$ will however
also shorten $x_0$. Preserving adiabaticity usually requires thus
a strong control pulse and an efficient storage will take place
when
\begin{equation}
x_{\max}\le x_0.
\end{equation}
Through their asymmetry in $\Omega_{p,\max}^2$ and
$\Omega_{c,\max}^2$ with respect to $q$, Eqs. (\ref{xmax}) and
(\ref{x0}) show that we can shorten $x_{\max}$ with the use of a
\textit{weak control} pulse in addition to using a small $q$
parameter, while preserving the condition of adiabaticity using a
\textit{strong probe} (see Fig. 2 for the pulse scheme).
\begin{figure}[hbtp]
 \centering
 \includegraphics[scale=0.8] {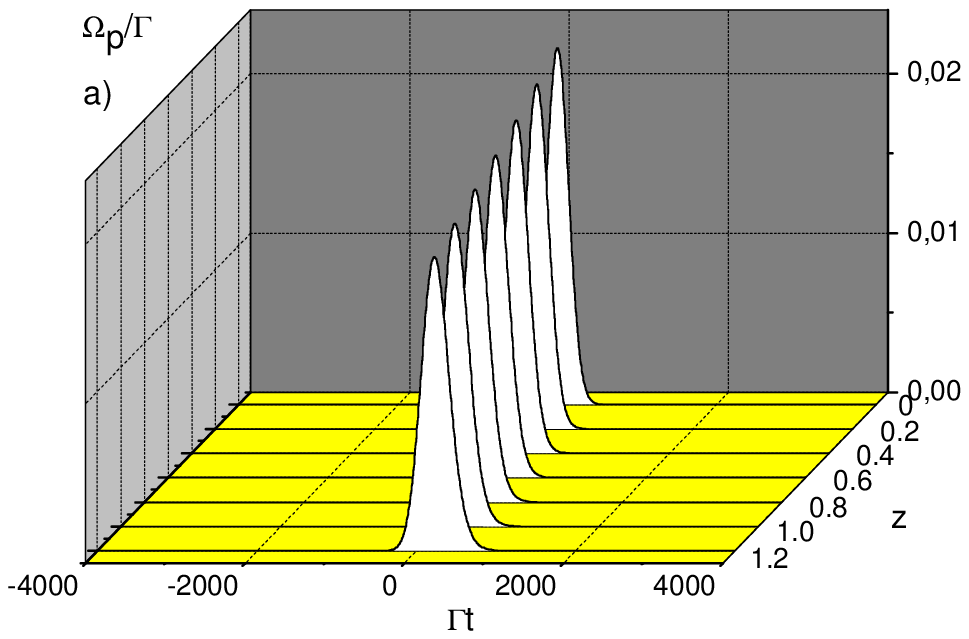}
 \includegraphics[scale=0.8] {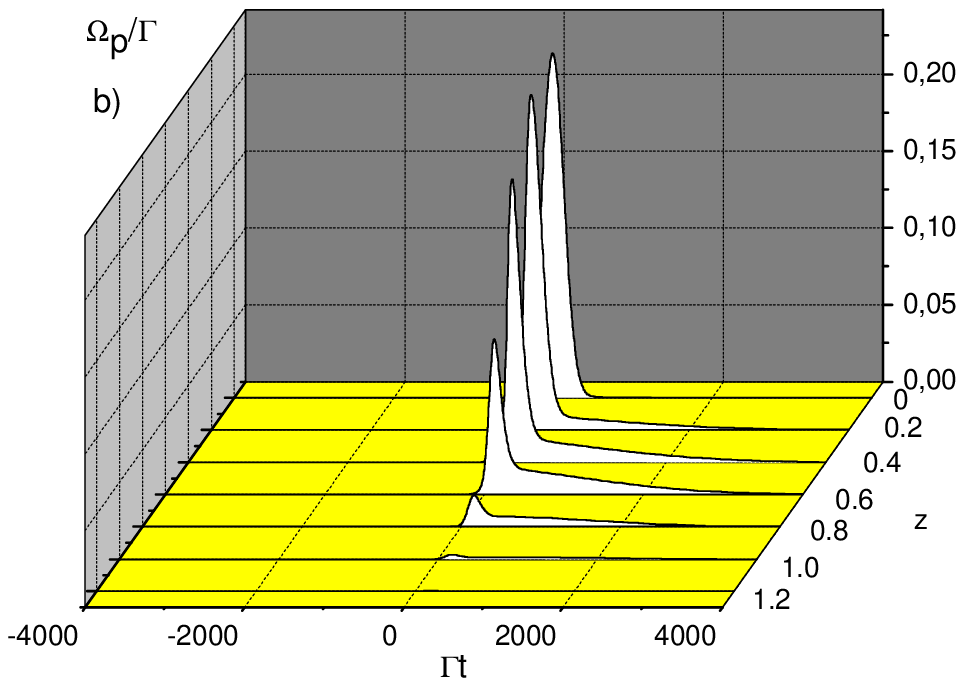}
 \caption{Propagation of the probe pulse for different propagation lengths
 $z$ for $\Delta=0$: a) Usual scheme of pulses with
Gaussian envelopes at the entrance of the medium displayed in Fig.
2a with $q=0.1$; b) Novel scheme of pulses (displayed in Fig.
2b).}
 \end{figure}
\\
 We have performed numerical calculations of the full
set of the density-matrix and Maxwell equations for both pulses
taking into account a fast relaxation $\Gamma T_p=200$. It indeed
shows that the storage of the probe field in inhomogeneously
broadened media is more efficient when we use atoms of different
adjacent transition strengths.
 Fig. 3b presents the temporal evolution of the probe
 pulse at different propagation lengths (normalized as
 $z\equiv xq_{p}/\Gamma$)
 for the novel scheme of the pulses at the entrance of the medium shown
 in Fig. 2b. The chosen parameters lead to
$\Omega_{c,\text{max}}T_{c}=\Omega_{p,\text{max}}T_{p} \approx
44$, $\Omega_{c,\text{max}}^2T_{c}/\Gamma \approx 1$, and
$\Omega_{p,\text{max}}^2T_{p}/\Gamma \approx 10$, which satisfy
the adiabatic conditions (\ref{conds}) and (\ref{cond_fast_relax})
at the entrance of the medium (here the time of interaction is
$T\sim T_p$). For the comparison we present in Fig. 3a the
dynamics for the usual scheme of pulses used for the optical
information storage. One can see from Figs. 3a and 3b that in the
case of the novel scheme the length of information storage is
dramatically reduced as compared with that of the usual case.
Namely, for the novel scheme at the propagation length $z$ of
about  1 the probe pulse is completely absorbed by the medium
while at the same length in the usual scheme the probe pulse
remains still unabsorbed.

 \section{Conclusion}
We have shown that the optical information storage of a probe in
inhomogeneously broadened media of $\Lambda$ atoms occurs for
short medium length when strong probe pulses, weak control pulses
and very different strengths of adjacent transitions are used. For
such a scheme the optical length may be of the order of unity in
contrast to the conventional scheme where the optical length must
well exceed unity. The proposed scheme would allow one to reduce
the unavoidable losses caused by decoherence. The required
$\Lambda$ systems with very different strengths of adjacent
transitions are expected to be found in, e.g., rare-earth doped
solid-state materials. Study to elaborate efficient schemes to
retrieve the stored pulse in this strong regime is in progress.


\begin{appendix}
\section{Characteristics method}
\label{app_characteristics} We briefly recall here the
characteristics method (see for instance \cite{Courant}) to solve
the Maxwell equation
\begin{equation}
\label{Maxwell_gen} \frac{\partial\theta}{\partial
x}+\frac{1}{u}\frac{\partial\theta}{\partial t}=0, \quad
\theta(0,t)\equiv\theta_0(t)
\end{equation}
in the case of a non-linear group velocity of the form: $u\equiv
u(x,t)$. We want to transform this linear first order partial
differential equation into an ordinary differential equation along
the characteristic curves $(x(s),t(s))$. Using the chain rule
$\frac{d}{ds}\theta(x(s),t(s))=\frac{\partial \theta}{\partial
x}\frac{dx}{ds}+\frac{\partial \theta}{\partial t}\frac{dt}{ds}$,
we obtain
\begin{equation}
\label{ODE} \frac{d}{ds}\theta(x(s),t(s))=0,
\end{equation}
if we set (choosing $x(0)=0$ and denoting $\xi:= t(0)$)
\begin{subequations}
\label{cond_ODE}
\begin{eqnarray}
\label{cond_ODEa} \frac{dx}{ds}&=&1,\text{ i.e. }x(s)=s,\\
\label{cond_ODEb} \frac{dt}{ds}&=&\frac{1}{u}, \text{ i.e. }
t(s)=\int_0^{s}\frac{ds'}{u(x(s'),t(s'))}+\xi.\quad
\end{eqnarray}
\end{subequations}
Equation (\ref{ODE}) leads to the solution
\begin{equation}
\label{ODE_sol}
\theta(x(s),t(s))=\theta(x(0),t(0))=\theta(0,\xi)=\theta_0(\xi)
\end{equation}
which is thus a constant along the characteristics.
$\xi$ allows one to label a characteristics $t(x)$ rewritten from
Eq. (\ref{cond_ODEb}) as (since $s=x$)
\begin{equation}
\label{cond_ODEc} t(x)=\int_0^{x}\frac{dx'}{u(x',t(x'))}+\xi.
\end{equation}
The non-linear velocity $u$ derived from (\ref{cond_ODEc}):
\begin{equation}
\label{nlv} \frac{1}{u}=\frac{dt}{dx}
\end{equation}
corresponds to the instantaneous velocity along a characteristic.

When $u$ is independent of $t$, i.e. $u\equiv u(x)$, one has
$\xi=t-\int_0^{x}dx'/u(x')$, which becomes $\xi=t-x/u$ if $u$ is
also independent of $x$.

When $u$ depends only on $t$, i.e. $u\equiv u(t)$, Eq.
(\ref{cond_ODEb}) can be rewritten as
\begin{equation}
\label{xi} \int_{\xi}^{t}dt'\,u(t')=s=x,
\end{equation}
which gives in principle $\xi\equiv\xi(x,t)$.

\end{appendix}

\vfill

\begin{acknowledgments}
We acknowledge support from INTAS 06-100001-9234, the Research
Project ANSEF PS-opt-1347 of the Republic of Armenia, the Agence
Nationale de la Recherche (ANR CoMoC) and the Conseil R\'{e}gional
de Bourgogne.
\end{acknowledgments}


\begin{references}

\bibitem{Lukin2003} M.D. Lukin, Rev. Mod. Phys. {\bf 75}, 457 (2003);
M. Fleischhauer, A. Imamoglu, J.P. Marangos, Rev. Mod. Phys. {\bf
77}, 633 (2005).

\bibitem{Liu}
C. Liu, Z. Dutton, C.H. Behroozi, L.V. Hau, Nature  {\bf 409},
490 (2001); T. Chaneliere, D.N. Matsukevich, S.D. Jenkins, S.Y. Lan,
T.A.B. Kennedy, A. Kuzmich. Nature {\bf 438}, 833, (2005).

\bibitem{Lukin2001}
M.D. Lukin, D.F. Phillips, A. Fleischhauer, A. Mair, R.L.
Walsworth, Phys. Rev. Lett. {\bf 86}, 783 (2001); A.S. Zibrov, A.B. Matsko, O. Kocharovskaya,
Y.V. Rostovtsev, G.R. Welch, M.O. Scully, Phys.Rev.Lett. {\bf 88}, 103601 (2002); M.D. Eisaman, A. Andre,
F. Massou, M. Fleischhauer, A.S. Zibrov, M.D. Lukin, Nature {\bf 438}, 837, (2005); R. Pugatch, M. Shuker,
O. Firstenberg, A. Ron, N. Davidson, Phys. Rev. Lett. {\bf 98}, 203601 (2007); P. K. Vudyasetu, R. M. Camacho,
J. C. Howell Phys. Rev. Lett. {\bf 100}, 123903 (2008).

\bibitem{Kuznetsova} E. Kuznetsova, O. Kocharovskaya, Ph. Hemmer,
M.O. Scully, Phys. Rev. A {\bf 66}, 063802 (2002); A.V. Turukhin,
V.S. Sudarshanam, M.S. Shahriar, J.A. Musser, B.S. Ham,  P.R. Hemmer.
Phys. Rev. Lett., {\bf 88}, 023602 (2002);  S.E. Yellin, P.R. Hemmer,
Phys. Rev. A, {\bf 66},
013803 (2002); L. Alexander, J. J. Longdell, M. J. Sellars, and N. B.Manson,
Phys. Rev. Lett. {\bf 96}, 043602 (2006); M. U. Staudt, S. R. Hastings-Simon,
M. Nilsson, M. Afzelius, V. Scarani, R. Ricken, H. Suche, W. Sohler,
W. Tittel, and N. Gisin, Phys. Rev. Lett. {\bf 98}, 113601 (2007); S$\o$rensen,
M. D. Lukin, and R. L. Walsworth, Phys. Rev. Lett. {\bf98}, 243602 (2007).

\bibitem{Johnsson} M. Johnsson, K. M$\o$lmer, Phys. Rev. A {\bf 70},
032320 (2004).

\bibitem {Novikova} I. Novikova, A. V. Gorshkov, D. F. Phillips,
A. S. S$\o$rensen, M. D. Lukin, R. L. Walsworth, Phys. Rev. Lett. {\bf 98},
243602 (2007); A. V. Gorshkov, A. Andr\'e, M. Fleischhauer,
A. S. S$\o$rensen, M. D. Lukin, Phys. Rev. Lett. {\bf 98}, 123601 (2007).

\bibitem {Shahriar}
7.  M.S. Shahriar, P.R. Hemmer, S. Lloyd, P.S. Bhatia, A. Craig,
Phys. Rev. A, {\bf 66}, 032301 (2002); M. Nilsson, L. Rippe, S. Kroll,
R. Klieber, D. Sutter, Phys. Rev. B, {\bf 70}, 214116 (2004).

\bibitem {Goto} 8.  J. Klein, F. Beil, T. Halfman, Phys. Rev. Lett. {\bf 99},
113003 (2007); H. Goto and K. Ichimura, Phys. Rev. A {\bf 75},
033404 (2007).

\bibitem{Dey} T.N. Dey , G.S. Agarwal, Phys. Rev. A {\bf 67}, 033813 (2003).

\bibitem{Grigoryan} G.G. Grigoryan, Y.T. Pashayan, Phys. Rev. A
{\bf 64}, 013816 (2001); V.O. Chaltykyan, G.G. Grigoryan, G.V. Nikoghosyan.
Phys. Rev. A, {\bf 68}, 013819, (2003).


\bibitem{Matsko} 14. A.B. Matsko, Y.V. Rostovtsev, O. Kocharovskaya, A.S. Zibrov,
M.O. Scully, Phys. Rev. A {\bf 64}, 043809 (2001).

\bibitem{Grobe} R. Grobe, F.T. Hioe, and J.H. Eberly, Phys. Rev. Lett.
\textbf{73}, 3183 (1994).

\bibitem{Courant} R. Courant, \textit{Partial Differential
Equations} (New York, 1962).

\end{references}
\end{document}